\documentclass[aps,pre,10pt,twocolumn,groupedaddress,showpacs]{revtex4-1}
\usepackage{amsmath,amsfonts,amssymb,bm,color,dsfont,graphicx,psfrag,mathtools}

\begin{document}

\title{Quasi-Critical Brain Dynamics on a Non-Equilibrium Widom Line}

\author{Rashid V. Williams-Garc\'{i}a}
\email[Electronic address: ]{rwgarcia@indiana.edu}
\author{Mark Moore}
\author{John M. Beggs}
\author{Gerardo Ortiz}

\affiliation{Department of Physics, Indiana University, Bloomington, Indiana 47405, USA}

\date{March 29, 2014}

\begin{abstract}
Is the brain really operating at a critical point? We study the non-equilibrium properties of a neural network which models the dynamics of the neocortex and argue for optimal quasi-critical dynamics on the Widom line where the correlation length and information transmission are optimized. We simulate the network and introduce an analytical mean-field approximation, characterize the non-equilibrium phase transitions, and present a non-equilibrium phase diagram, which shows that in addition to an ordered and disordered phase, the system exhibits a quasiperiodic phase corresponding to synchronous activity in simulations which may be related to the pathological synchronization associated with epilepsy.
\end{abstract}

\maketitle

\section{Introduction}
Recent experimental evidence from a variety of living neural networks suggests that the brain may be operating at or near a critical point, poised between disordered (``subcritical'') and ordered (``supercritical'') phases where cascades of activity are damped or amplified, respectively \cite{BeggsPlenz2003,Pasquale2008,Petermann2009,Kitzbichler2009,Poil2008,Shriki2013,Ribeiro2010,Priesemann2013,Friedman2012}. At this interface, neural networks are expected to produce avalanches of activity whose size and duration probability distributions follow power laws, as a distinctive feature of critical phenomena is scale-invariance \cite{Levina2007,HaldemanBeggs2005,Harris1989,NishimoriOrtiz2011}. Theory and simulations conjectured that neural networks poised at a critical point would have optimal information transmission \cite{BeggsPlenz2003}, information storage \cite{HaldemanBeggs2005,Chen2010}, computational power \cite{BertschingerNatschlager2004,Gollo2013}, dynamic range \cite{KinouchiCopelli2006,Larremore2011,Publio2012,Manchanda2013,Mosqueiro2013,Wang2013,Gollo2013,Pei2012}, and learning capabilities \cite{deArcangelisHerrmann2010}, while providing flexible, yet stable dynamics \cite{HaldemanBeggs2005,Magnasco2009}. Several experiments claim results consistent with these predictions \cite{Shew2009,Shew2011,Solovey2012}, lending plausibility to the criticality hypothesis of brain function \cite{Beggs2008}.

Here we introduce and analyze the so-called {\it cortical branching model} (CBM), a non-equilibrium stochastic cellular automaton capturing many features of neural network data \cite{HaldemanBeggs2005,Chen2010,PajevicPlenz2009}, and develop an analytical mean-field approximation in the form of an autonomous nonlinear discrete dynamical map of first order and dimension given by the integer-valued refractory period. We establish the non-equilibrium phase diagram of the CBM and identify three separate phases: the disordered, the ordered, and the quasiperiodic phases. Using this mean-field approximation, we argue that a continuous phase transition between the disordered and ordered phases occurs (in the thermodynamic limit) only when external driving, which we model as the spontaneous activation of network elements, is absent.

In our CBM, when external driving is present (a key feature of open dynamical systems), we find that this phase transition disappears and hence argue that true criticality is not attainable by living neural networks. We thus introduce an extension, along with a more proper quantitative formulation of the \emph{quasi}-criticality hypothesis. Our quasi-criticality hypothesis involves a non-equilibrium Widom line of maximum (though finite) dynamical susceptibility along which correlation length and, as we shall demonstrate, mutual information are maximized. We expect that quasi-critical behavior can be observed along this line: for instance, distributions of activity avalanches are nearly power-law and avalanche shape collapses can be approximately performed to yield approximate scaling exponents \cite{Papanikolaou2011,Friedman2012}. Moreover, this Widom line framework quantifies the notion of proximity of our neural system to its unattainable non-equilibrium critical point, i.e. we now know how to drive the system towards or away from its optimal behavior, by manipulating the relevant parameters.

Additionally, increasing the refractory period at large values of the branching parameter, induces a quasiperiodic phase in the mean-field which corresponds to synchronous activation in simulations. Results of our numerical simulations are qualitatively consistent with the mean-field calculations as long as the graph underlying the complex network is irreducible. Because spontaneous activation rates in neural networks are readily manipulated experimentally \cite{Mazzoni2007,Gunning2013}, our predictions could soon be tested; it is worth noting that our approach can be extended to other systems, such as the SIRS compartmental disease epidemic model, which shares many similarities with the CBM \cite{AndersonMay1979}, although the latter is more general. In our concluding remarks, we describe how to experimentally control various parameters involved in the CBM to assess the validity of the quasi-criticality hypothesis.

\section{The Cortical Branching Model}
\label{sec:theCBM}
\begin{figure}[htp]
	\includegraphics[width=\columnwidth]{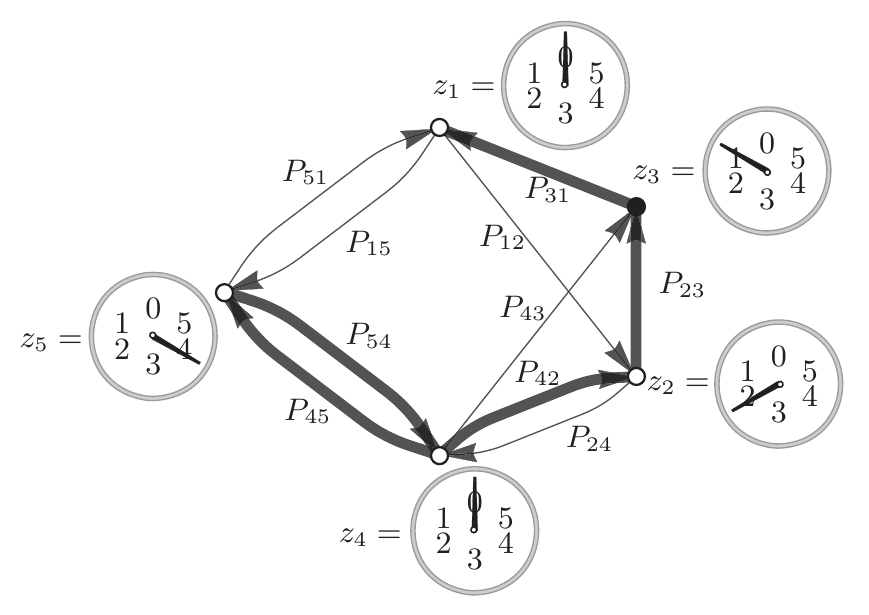}
	\caption{A random, directed network of $N=5$ nodes (vertices). Each node has $k_{\sf in}=2$ incoming connections (edges), each of which are weighted; the thickness of the edges illustrate the connection strengths $P_{ij}$. Node $3$ is active ($z_3=1$); nodes $1$ and $4$ are quiescent ($z_i=0$ for $i=1,4$); and nodes $2$ and $5$ are refractory.}
	\label{AdjacencyPlot}
\end{figure}

We next introduce details of the CBM. Consider a random directed network, or graph, of $N$ nodes, where each node has its own local neighborhood of interactions; connections are established and kept fixed throughout the dynamics, as in quenched disorder. Random networks can either be \emph{strongly-connected}--in which case there exists a path (never running anti-parallel through directed connections) from any node in the network to any other node on the network (through possibly many intermediaries)--or \emph{weakly-connected}--in which case the network contains disjoint subgraphs and is said to not be fully-connected. Networks are generated randomly and tested for connectedness by examining the corresponding adjacency matrix associated with its graph. In this study, we only consider strongly-connected networks, i.e. those with \emph{irreducible} adjacency matrices \cite{Larremore2012}. See Fig. \ref{AdjacencyPlot} for a sample network.

Internodal connections are weighted, with elements of the weighted adjacency matrix $P=\{P_{ij}\leq1\}$ representing the probability $P_{ij}=\kappa p_{n_{ij}}$ that a connection from node $i$ to node $j$ will transmit activity, with
\begin{equation} \label{Pij}
p_{n_{ij}} = \frac{e^{-B n_{ij}}}{\sum_{n=1}^{k_{\sf in}} e^{-B n}} ,
\end{equation}
where $\kappa$ is the branching parameter (which is equivalent to the Perron-Frobenius eigenvalue of $P$), $k_{\sf in}$ is the in-degree of each node, $B$ is the connection strength bias, and $n_{ij} \in \{1,\cdots,k_{\sf in}\}$ ranks each connection inbound at node $j$ by strength, e.g. $n_{ij}=1$ corresponds to the strongest connection inbound at node $j$. We restrict $\kappa$ to the range $\lbrack0,\kappa_{\sf max}\rbrack$, where the upper bound is given by $\kappa_{\sf max}=e^B\sum_{n=1}^{k_{\sf in}}e^{-Bn}$ and the lower bound corresponds to a fully-disconnected network. Close to and above $\kappa=\kappa_{\sf max}$, the CBM produces constant activity, i.e. $\rho_1(t)\neq0$ for all times $t$ (a single avalanche of infinite duration). It had previously been determined that for a network of $N=60$ nodes, each with a fixed $k_{\sf in}=10$, that the values $B=1.2$ and $B=1.6$ allowed for a reasonable fit to the local field potential (LFP) dynamics recorded from living neural networks \cite{Chen2010}; we present our primary simulation results with $B=1.4$ and $k_{\sf in}=3$.

The state of each node $i$ is described by a dynamical state variable $z_i\in S$, where $S=\{0,1,2,\hdots,\tau_{\sf r}\}$, $i=1,\cdots,N$, and $\tau_{\sf r}\geq1$ is the integer-valued refractory period, i.e. the number of time steps following \emph{activation} during which a node cannot be made to activate. We define the configuration space of the CBM as ${\cal C}=\{Z=(z_1,z_2,\hdots,z_N)|z_i \in S\}_{i=1,N}$, where $\dim{\cal C}=(\tau_{\sf r}+1)^N$; for example, ${\cal C}=\{(0,0);(0,1);(1,0);(1,1)\}$ for a system of $N=2$ and $\tau_{\sf r}=1$. A node $i$ is said to be \emph{active} when $z_i=1$, \emph{inactive} (i.e. quiescent) when $z_i=0$, and \emph{refractory} at any other value. Nodes can only be active for a single time step at a time.

The system is driven by the spontaneous activation of a node, which occurs with probability $p_{\sf s}$. The number of time steps between spontaneous activations follows a discrete probability distribution of our choice: a Poisson distribution with rate $1/(p_{\sf s}N)$, i.e. $P(\Delta t_s) = (p_{\sf s}N)^{-\Delta t_s}e^{-1/p_{\sf s}N}/{\Delta t_s!}$, allows for a greater separation of driving and relaxation timescales, such as that seen in instances of self-organized criticality (SOC) \cite{Jensen1998,VespignaniZapperi1998}, thus minimizing the occurrence of overlapping avalanches; whereas by using a geometric distribution with success probability $p_{\sf s}N$, i.e. $P(\Delta t_s) = (1-p_{\sf s}N)^{\Delta t_s-1}p_{\sf s}N$, avalanches are more likely to overlap and contain spontaneous events. Simulation results presented herein utilize Poisson-distributed spontaneous events to generate avalanches.

A node can also be driven to activate by another node connected to it with probabilities given by Eq. \eqref{Pij}, but only if the driving node was active and the driven node quiescent in the preceding time step. Regardless of the method of stochastic activation, a node's dynamical variable $z_i$ changes deterministically following activation, increasing by $1$ every time step until $z_i=\tau_{\sf r}$ is reached, after which the node becomes quiescent ($z_i=0$) until it is stochastically activated once again. Thus, each state variable $z_i$ represents a {\it clock} degree of freedom. For example, consider a node $i$ with $\tau_{\sf r}=3$: following the time step during which it was active, this node will become refractory, its state deterministically changing from $z_i=2$ to $z_i=3$, and finally to $z_i=0$.

We summarize the dynamics of the random neighbor discrete CBM with the following algorithm:
\newcounter{lcounterRN}
\begin{list}{\bf{\arabic{lcounterRN}}.~}{\usecounter{lcounterRN}}
	\item \emph{Initialization.} Prepare nearest neighbor connections by randomly assigning connections between nodes while keeping the in-degree $k_{\sf in}$ fixed (parallel connections are allowed; loops are not) and prepare connection strengths $P_{ij}$ as given by Eq. \eqref{Pij}. Initialize the system in the only stable configuration, i.e. $z_i = 0$ for every node $i$. Prepare the first spontaneous activation(s) at $t=1$ and subsequent spontaneous activation times by drawing inter-activation intervals $\Delta t_s$ from a Poisson distribution.
	
	\item \emph{Drive.} For each spontaneous activation time equal to the current time step $t$, randomly select a node $j$ to activate, $z_j(t) \rightarrow 1$; if however node $j$ was not initially quiescent (i.e. $z_j(t)=0$), then spontaneous activation does not occur at node $j$.
	
	\item \emph{Relaxation.} Any nodes $i$ for which $z_i(t-1) \neq 0$: $z_i(t) = z_i(t-1)+1$. If $z_i(t) > \tau_{\sf r}$, then $z_i(t) \rightarrow 0$. Node $j$, having been active at time step $t$, will influence the activity of a neighboring node $k$ at time step $t+1$ with probability $P_{jk}$, but only if $z_k(t+1)=0$: $z_k(t+1) \to z_k(t+1)+1$.
	
	\item \emph{Iteration.} Start the next time step: Return to \bf{2}.
\end{list}

\section{Avalanche Characterization}
Spatio-temporal clusters of activation (avalanches) exhibited by the CBM mimic spatio-temporal patterns (\emph{neuronal} avalanches) observed in living neural networks \cite{HaldemanBeggs2005,Chen2010}. We explore their properties by first defining the density of active nodes at time $t$, $\rho_1(t)$, as
\begin{equation}\label{rho1}
	\rho_1(t) = \frac{1}{N}\sum_{i=1}^N \delta_{z_i(t),1},
\end{equation}
although we often consider its time average, $\bar{\rho}_1=\langle \rho_1(t) \rangle_t = 1/N_{\sf T} \sum^{N_{\sf T}}_{t=1}\rho_1(t)$, where $N_{\sf T}$ is the total number of time steps. The \emph{zero-field} dynamical susceptibility $\chi$, associated with the density of active nodes,  corresponds to the fluctuation of $\rho_1(t)$, $\chi=N[\langle\rho_1^2(t)\rangle_t-(\bar{\rho}_1)^2]$, and quantifies the dynamical response of the system. The correlation length associated with $\chi$ will play an important role in establishing the quasi-criticality hypothesis. In the mean-field approximation defined below, $\chi$ will be determined from the expression $\lim_{p_{\sf s}\rightarrow0} \partial \bar{\rho}_1 / \partial p_{\sf s}$.

Periods of inactivity ($\rho_1=0$) are punctuated by periods of activity ($\rho_1\neq0$) which constitute avalanches. The properties of these avalanches are encoded in the avalanche \emph{shape}, which we define as the density of active nodes over the duration of an avalanche, resembling definitions given in previous studies \cite{SethnaDahmenMyers2001}. The avalanche shape vector $X_q$ gives the shape of the $q$th avalanche:
\begin{equation}
	X_q(\phi) = \sum_{i=1}^N \delta_{z_i(t^0_q+\phi-1),1} ,
\end{equation}
where $t^0_q$ is its starting time, $d_q$ is its duration, and $\phi=[1,d_q] \in \mathbb{Z}^+$ indexes the number of time steps within the avalanche. From this, we write the size of the $q$th avalanche as $s_q=\sum_{\phi=1}^{d_q} X_q(\phi)$.

Avalanche size and duration probability distributions are conjectured \cite{BeggsPlenz2003} to follow power laws, $P(s) \propto s^{-\tau}$ and $P(d) \propto d^{-\alpha}$. In simulated and living neural networks, values of these exponents have been found to be $\tau\approx1.5$ and $\alpha\approx2$ for LFP data and $\tau\approx1.6$ and $\alpha\approx1.7$ for neuronal spike data; results which have been used to support the criticality hypothesis \cite{Levina2007,Harris1989,HaldemanBeggs2005}.

\section{A Mean-Field Approximation}
In order to gain a deeper understanding of our CBM and its non-equilibrium phase diagram, we next develop an analytical mean-field approximation. In the mean-field approximation, a typical, \emph{representative} node and its local neighborhood of interaction (i.e. the $k_{\sf in}$ sites which directly influence its behavior) are used to approximate the behavior of the network as a whole--the key presumption here being that transition probabilities are translationally invariant in the thermodynamic limit and beyond the upper critical dimension. We would expect the mean-field approach to represent a faithful approximation of the simulation results when the simulated graph is irreducible; it is an extremely interesting question to explore the cases where the graph is reducible, but this is beyond the scope of the current paper. The cellular automaton rules of the CBM (described above in Section \ref{sec:theCBM}) are approximated by a Markovian stochastic process and so the probability that a particular node will be in a specific state is given by the Chapman-Kolmogorov equation \cite{DeutschDormann2004}:
\begin{equation}\label{Chap-Kol}
 	P(z_r(t+1)=z) = \sum_{\textbf{z} \in S^{k_{\sf in}+1}}W(\textbf{z} \rightarrow z)\prod_{i=0}^{k_{\sf in}}P(z_i(t)),
\end{equation}
where $z$ is an element in the state space $S=\{0,...,\tau_{\sf r}\}$, $r\in\{0,...,k_{\sf in}\}$ identifies the nodes (with $r=0$ corresponding to the representative node), $\textbf{z}=(z_0,...,z_{k_{\sf in}})$ is the configuration of the system (i.e. a vector whose elements are the states of the representative node and its local neighborhood of interaction), and $W(\textbf{z} \rightarrow z)$ is the probability that the $r=0$ node will transition into state $z$ given the system configuration $\textbf{z}$. At a particular iteration of the mean-field, $t$, the probability that a node $r$ is in state $z$ is equivalent to the fraction of nodes $x_z(t)$ in state $z$: $P(z_r(t)=z) = x_z(t)=\sum_{i=0}^{k_{\sf in}} \delta_{z_i(t),z}/(k_{\sf in}+1)$. Additionally, because we are primarily interested in the density of active nodes $x_1$ and because a node must be quiescent at $t$ to become active at $t+1$, we rewrite Eq. \eqref{Chap-Kol} as
\begin{equation}\label{MFmap}
 	x_1(t+1)=x_0(t)\sum_{\textbf{z}^\prime \in S^{k_{\sf in}}}W(\textbf{z}^\prime \rightarrow 1)\prod_{j=1}^{k_{\sf in}}x_{z_j}(t),
\end{equation}
where $\textbf{z}^\prime$ is the configuration of the local neighborhood \emph{excluding} the representative node, i.e. $\textbf{z}^\prime=(z_1,...,z_{k_{\sf in}})$. We write a general expression for the transition probabilities $W(\textbf{z}^\prime \rightarrow 1)$ as one minus the probability that a node will remain quiescent, or
\begin{equation}
 	W(\textbf{z}^\prime \rightarrow 1) = 1-(1-p_{\sf s})\prod_{j=1}^{k_{\sf in}}(1-\kappa p_j\delta_{{z_j},1}),
\end{equation}
where the connection strengths $p_j$ are of the form given by Eq. \eqref{Pij}. Because $z$ varies deterministically following activation, $x_z(t+1)=x_{z-1}(t)$ for $z\in\{2,...,\tau_{\sf r}\}$.

Along with Eq. \eqref{MFmap}, these equations form a nonlinear, autonomous $(\tau_{\sf r}+1)$-dimensional map of first order (i.e. Markovian). By including the restriction that, at any iteration $t$, $\sum_{z=0}^{\tau_{\sf r}}x_z(t)=1$, we reduce the dimension to $\tau_{\sf r}$. This map then allows us to calculate the mean-field densities of quiescent ($z=0$), active ($z=1$), and refractory nodes. An equivalent mean-field approximation can be formulated as a non-Markovian $\tau_{\sf r}$th-order map in one dimension. Finally, we note that increasing the refractory period by a single time step increases the number of equations by one; whereas increasing $k_{\sf in}$ increases the order of polynomial to be solved. Fixed points $x_1^*$ of this map give approximate densities of active sites, i.e. mean-field approximations to Eq. \eqref{rho1}. Stability of each fixed point is determined as usual by calculating the eigenvalues of the Jacobian matrix associated with the map; if each of the eigenvalues of the Jacobian when evaluated at a certain fixed point have modulus less than one, then that fixed point is stable.

\subsection{Non-Equilibrium Phase Diagram and the Widom Line}
We first consider the case $k_{\sf in}=1$. The mean-field approximation in this case is given by the quadratic map
\begin{align}\label{kin1MF} \nonumber
  x_1(t+1)=&{}\left(1-\sum_{z=1}^{\tau_{\sf r}}x_z(t)\right) \lbrack c \, x_1(t)+p_{\sf s} \rbrack \\
  x_z(t+1)=&\:x_{z-1}(t) \text{, for } z=\{2,\cdots,\tau_{\sf r}\} ,
\end{align}
where $c=\kappa p_1 (1-p_{\sf s})$. This yields two fixed points, which when $p_{\sf s}=0$ are $x_1^*=0$ and $x_1^*=(1-1/\kappa p_1)/\tau_{\sf r}$. The vanishing fixed point becomes unstable when $\kappa>1$ and so the stable fixed point acts as a Landau order parameter, i.e. $\bar{\rho}_1=0$ for $\kappa \leq 1$ and $\bar{\rho}_1>0$ for $\kappa > 1$, with the critical point at $\kappa_{\sf c}=1$. We find the critical exponent $\beta=1$: $x_1^*\propto(\kappa-\kappa_{\sf c})^\beta$ for $\kappa>1$. Calculating the susceptibility, $\chi=\lim_{p_{\sf s}\rightarrow0} \partial \bar{\rho}_1 / \partial p_{\sf s}$, we find that it diverges at $\kappa_{\sf c}$ with exponent $\gamma'=1$ for $\kappa<1$: $\chi \propto(\kappa_{\sf c}-\kappa)^{-\gamma'}$. For $\kappa>1$, it diverges with exponent $\gamma=1$: $\chi \propto(\kappa-\kappa_{\sf c})^{-\gamma}$.

It is remarkable to note that the $k_{\sf in}=1$ CBM mean-field approximation is the discrete-time equivalent of the {\it directed percolation} (DP) mean-field equation when $\tau_{\sf r}=1$:
\begin{equation} \label{DPMF}
  \partial_t\rho_1(t)=-c\rho_1(t)^2+(c-1-p_{\sf s})\rho_1(t)+p_{\sf s}
\end{equation}
as given in \cite{Henkel2008}, where $p_{\sf s}$ plays the role of an external field. These two seemingly different processes are therefore related even when $p_{\sf s}\neq0$. We note that the CBM has a continuous phase transition when $p_{\sf s}=0$, characterized by the order parameter $\bar{\rho}_1$, but that transition disappears when $p_{\sf s}\neq0$. The case $p_{\sf s}=0$ is consistent with the Janssen-Grassberger conjecture \cite{Janssen1981,Grassberger1982}, which states that a model with a continuous phase transition should belong to the DP universality class if the transition is characterized by a non-negative one-component order parameter.

In the case $k_{\sf in}=2$, the mean-field approximation yields the following cubic map:
\begin{align}\label{kin2MF} \nonumber
  x_1(t+1)=&{}\left(1-\sum_{z=1}^{\tau_{\sf r}}x_z(t)\right) \lbrack -ax^2_1(t)+bx_1(t)+p_{\sf s} \rbrack \\
  x_z(t+1)=&\:x_{z-1}(t) \text{, for } z=\{2,\cdots,\tau_{\sf r}\} ,
\end{align}
where $a=\kappa^2 p_1 p_2 (1-p_{\sf s})$ and $b=\kappa (1-p_{\sf s})$. In the absence of spontaneous activity, $p_{\sf s}=0$, we again have a vanishing fixed point, $x_1^*=0$, but now a pair of real, non-zero fixed points given by
\begin{equation}\nonumber
  x_{1\pm}^*=\frac{\kappa p_1 p_2+\tau_{\sf r}\pm\sqrt{(\kappa p_1 p_2+\tau_{\sf r})^2-4 p_1 
  p_2 \tau_{\sf r} (\kappa-1)}}{2 \kappa p_1 p_2 \tau_{\sf r}}.
\end{equation}
Expanding $x^*_{1-}$ around $\kappa = \kappa_{\sf c}$, we again find $x^*_{1-}\propto(\kappa-\kappa_{\sf c})^{\beta}$ with $\beta=1$. The zero-field dynamical susceptibility is then found to be
\begin{equation}
	\chi(\kappa)=\frac{f}{g(p_{\sf s}=0,\kappa)},
\end{equation}
where $f=1+(p_1p_2-1)x^*-p_1p_2x^{*3}$ and $g(p_{\sf s},\kappa)=(1-\kappa)+(1+\kappa)p_{\sf s}-2(\kappa+p_1p_2)(p_{\sf s}-1)x^*+3p_1p_2(p_{\sf s}-1)x^{*2}$, where $x^*$ is taken to be $x^*_1=0$ below the critical point ($\kappa<\kappa_{\sf c}$) and $x_{1-}^*$ above it ($\kappa>\kappa_{\sf c}$). Critical exponents of $\chi(\kappa)$ below and above the critical point are hence found to be $\gamma'=1$ and $\gamma=1$, respectively. Note that $\chi(\kappa)$ diverges at $\kappa_{\sf c}=1$ only when $p_{\sf s}=0$.

\begin{figure}[htp]
	\includegraphics[width=\columnwidth]{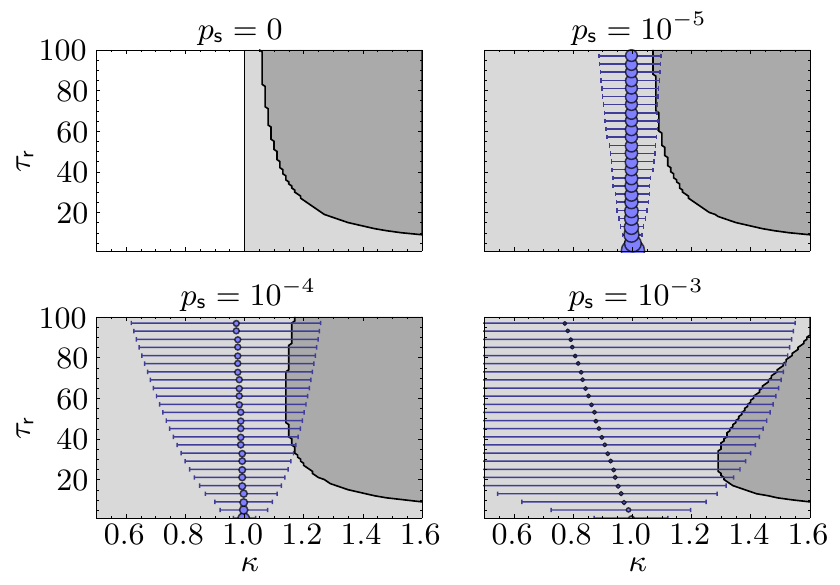}
	\caption{(Color online) Non-equilibrium mean-field phase diagram for $k_{\sf in}=2$, at selected values of $p_{\sf s}$.
	The white region corresponds to the subcritical disordered phase with a vanishing stable fixed point; the light-gray region corresponds to the supercritical ordered phase with a nonzero stable fixed point; the dark-gray region corresponds to an ``oscillatory'' quasiperiodic phase, where all fixed points are unstable. Solid black lines are lines of non-analyticity and thus represent phase boundaries.}
\label{PhaseDiag}
\end{figure}

Stability of the fixed points again changes at $\kappa_{\sf c}=1$. The fixed point $x^*_1=0$ is stable when $\kappa<\kappa_{\sf c}$ for any value of $\tau_{\sf r}$; this defines the \emph{disordered phase}. Stability shifts to the fixed point $x^*_{1-}$ when $\kappa>\kappa_{\sf c}$--defining the \emph{ordered phase}--but only for small values of $\tau_{\sf r}$. With $B=0.5$ and $\kappa=\kappa_{\sf max}$, all fixed points lose stability when $\tau_{\sf r}\geq9$, where $\kappa_{\sf max}\approx1.607$. Indeed this defines a new phase boundary which separates the ordered phase from an \emph{entirely different} phase, where the CBM exhibits quasi-periodic behavior (see Fig. \ref{PhaseDiag}, top left box).

When $p_{\sf s}\neq0$ (and for small values of $\tau_{\sf r}$), $x_{1-}^*$ is stable across $\kappa=1$ and the dynamical susceptibility $\chi$ no longer diverges, i.e. the phase transition disappears, giving way to a crossover region (see Fig. \ref{PhaseDiag}). To give an idea of the shape of $\chi$, we have included light-blue bubbles with diameter logarithmically-scaled to its magnitude, and blue horizontal lines indicating its width at half-maximum which encompasses the {\it quasi-critical} region. We have used the value $B=0.5$ for presentation purposes, as it allows for a better view of the extent of the quasiperiodic phase boundary when $\kappa$ is large; note that with $k_{\sf in}=2$, $\kappa_{\sf max}(B=1.4)\approx1.247$ while $\kappa_{\sf max}(B=0.5)\approx1.607$. Changes in $B$ had no discernible impact on the phase diagram. For a fixed value of $\tau_{\sf r}$, we can identify the peak in the susceptibility (and correlation length), defining a non-equilibrium Widom line in the $\tau_{\sf r}$--$\kappa$ plane; for the equivalent equilibrium Widom line see \cite{Xu2005}.

\subsection{The Quasiperiodic Phase}

\begin{figure}[htp]
	\includegraphics[width=\columnwidth]{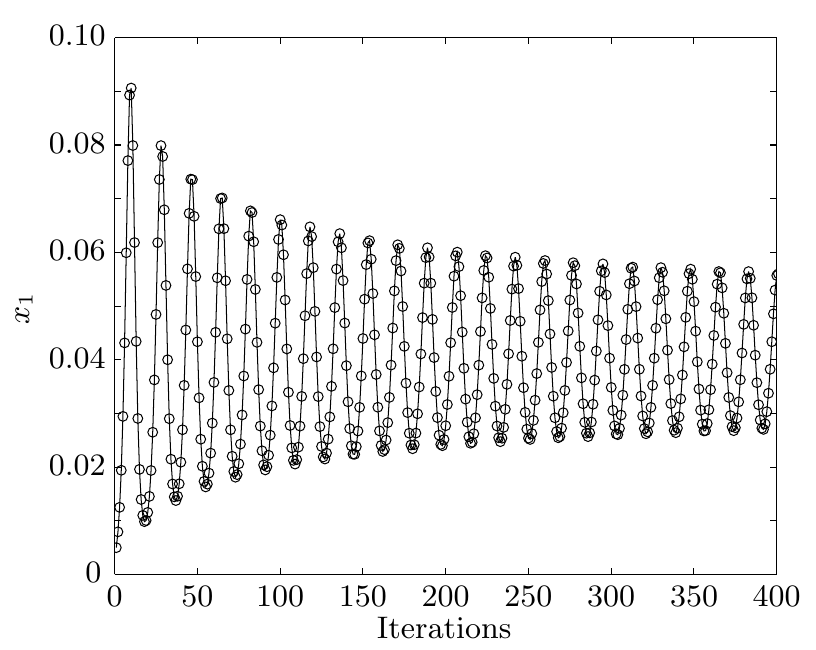}
	\caption{CBM mean-field density of active nodes over $400$ iterations showing quasi-periodic behavior; $k_{\sf in}=2$, $B=0.5$, $\kappa = 1.60$, $\tau_{\sf r}=9$, $p_{\sf s}=0$.}
	\label{kin2Osc}
\end{figure}

For large $\tau_{\sf r}$, all fixed points of the $k_{\sf in}=2$ mean-field lose stability and the mean-field density of active nodes $x_1(t)$ subsequently exhibits ``oscillatory'' behavior as presented in Fig. \ref{kin2Osc}; similar quasiperiodic phenomena had previously been observed in SIRS-like models \cite{Girvan2002}. Within this quasiperiodic phase, $x_1(t)$ does not converge to a fixed-point and periodic points are not present (hence ``quasiperiodic''). The envelope of $x_1(t)$ is, however, sinusoidal here (see Fig. \ref{kin2Osc}). This quasiperiodic phase was found to diminish with increasing $p_{\sf s}$, eventually disappearing at $p_{\sf s}\approx7 \times 10^{-3}$. Oscillatory behavior emerging at large refractory periods had previously been observed in neural network models \cite{RozenblitCopelli2011,CurtuErmentrout2001,KinouchiCopelli2006}, but the quasiperiodic behavior observed here and in \cite{Girvan2002} was not found. It should be noted that Curtu and Ermentrout's model involved both excitatory and inhibitory elements whereas the CBM only involves excitatory elements \cite{CurtuErmentrout2001}.

\section{Simulation of Avalanche Physics}
We now go beyond the mean-field and present results from CBM simulations which demonstrate the presence of a non-equilibrium Widom line, a non-equilibrium phase diagram qualitatively similar to the mean-field non-equilibrium phase diagram, and a quasiperiodic phase. Because the mean-field approximation eliminates the fluctuations responsible for avalanches, it is not useful in analyzing the statistics of the avalanches associated with our model and so we also utilize results from our CBM simulations to study the avalanche physics and prepare avalanche size distributions. We re-emphasize the use of irreducible graphs in simulating the CBM.

\begin{figure}[htp]
	\includegraphics[width=\columnwidth]{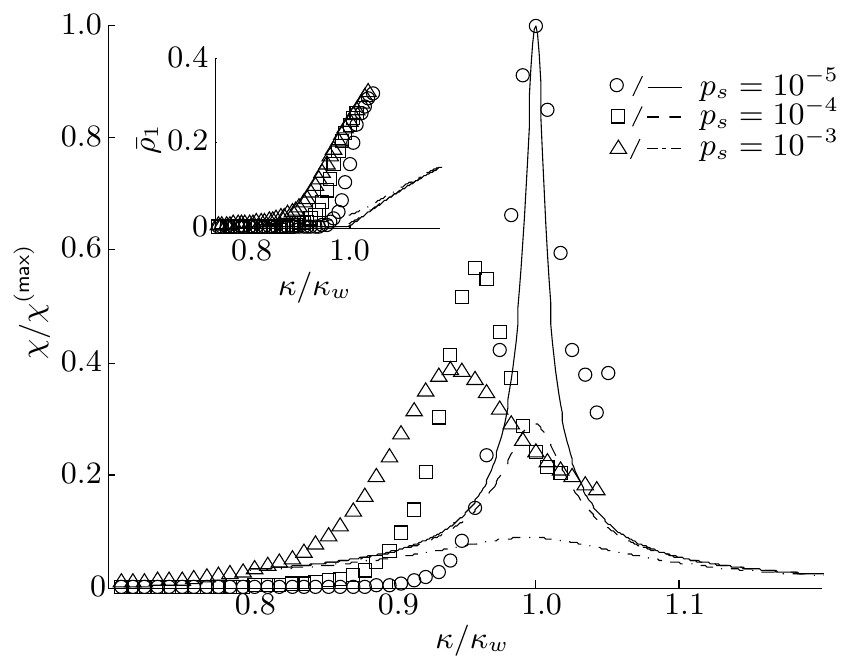}
	\caption{Dynamical susceptibility $\chi$ and $\bar{\rho}_1$ (inset) results from simulation with $N=128$ (data markers) and mean-field approximation (lines). Results are normalized to their maximal values and plotted against $\kappa$ normalized to the quasi-critical point $\kappa_{\sf w}$ at $p_{\sf s}=10^{-5}$. For simulations, we find $\kappa_{\sf w}$ to be $1.10$, $1.12$, and $1.17$ at $p_{\sf s}=10^{-3}$,  $10^{-4}$, and $10^{-5}$, respectively.}
	\label{DynamicalChi}
\end{figure}

\begin{figure}[htp]
	\includegraphics[width=\columnwidth]{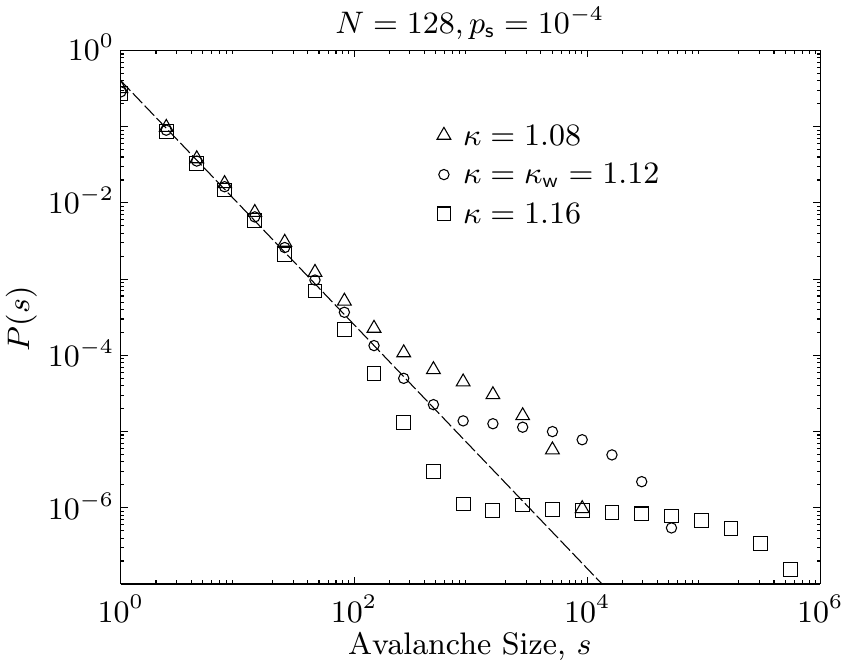}
	\caption{Logarithmically-binned avalanche size probability distributions $P(s)$ at various values of $\kappa$. The dashed line represents a power law with exponent $\tau=1.6$.}
	\label{PoS}
\end{figure}

We performed simulations using system sizes $N=\{32,64,96,128\}$ with $p_{\sf s}=\{10^{-5},10^{-4},10^{-3}\}$ at each value of $\kappa=\lbrack 0.8,1.3 \rbrack$ with a step size $\delta\kappa=0.01$ and with $\tau_{\sf r}=1$. Simulations were performed on ten different random networks until $10^6$ avalanches were generated at each value of $\kappa$; avalanche durations were limited to $10^5$ time steps. Note that with $B=1.4$ and $k_{\sf in}=3$, $\kappa_{\sf max} \approx 1.307$. We determined the time-averaged density of active nodes $\bar{\rho}_1$ as well as the dynamical susceptibility $\chi$, each as functions of $\kappa$ for the various values of $p_{\sf s}$ for simulations and mean-field for comparison. The dynamical susceptibility peaks at quasi-critical points $\kappa_{\sf w}$ defining a non-equilibrium Widom line in the $p_{\sf s}$--$\kappa$ plane (see Fig. \ref{DynamicalChi}). Avalanche size distributions at these $\kappa_{\sf w}$ exhibit quasi-power-law behavior over a maximum number of decades (see Fig. \ref{PoS}).

Much of the disagreement between the mean-field and simulation results is due to finite-size effects. If we were interested in the thermodynamic limit, however, we would need much larger system sizes which would require correspondingly large $k_{\sf in}=\eta N$ for $0<\eta\leq1$ such that the simulated networks maintained irreducibility; this quickly becomes numerically-intensive and computationally-complex.

\begin{figure}[htp]
	\includegraphics[width=\columnwidth]{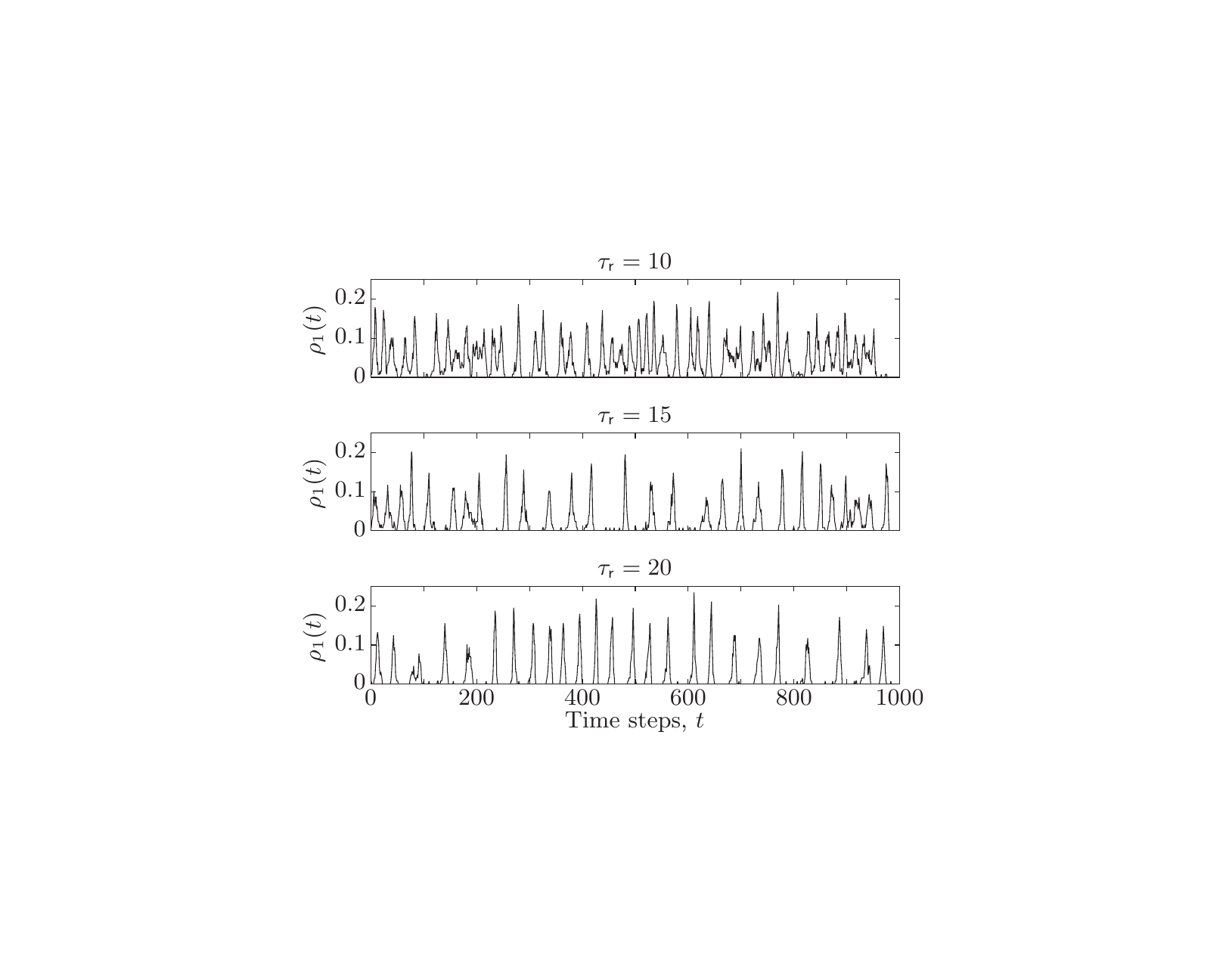}
	\caption{CBM simulation density of active nodes over $1000$ time steps; $N=128$, $k_{\sf in}=3$, $B=0.5$, $\kappa = 1.60$, and $p_{\sf s}=10^{-3}$.}
	\label{SimOsc}
\end{figure}

At values of $\tau_{\sf r}$ and $\kappa$ comparable to those at which the mean-field exhibits quasiperiodicity, an oscillatory synchronization phenomenon is observed in simulations (see Fig. \ref{SimOsc}). At high $\kappa$ and low $\tau_{\sf r}$, activity is nearly constant and very few avalanches are produced. As $\tau_{\sf r}$ is increased, large populations of nodes activate and become refractory long enough for avalanches to be produced once again. Note that avalanches produced under these conditions are not scale-free, since the typical avalanche size approaches the system size.

\section{Optimal Information Transmission and the Widom Line}

Mutual information has previously been used to measure information transmission in neural networks \cite{Zador1998} and to demonstrate that information transmission is optimized at, or in the vicinity of phase transitions \cite{Matsuda1996,BeggsPlenz2003,Wicks2007,Shew2011}. To investigate this in random networks of the CBM, we hence compute the mutual information $I_T({\sf S};{\sf R})$ from an ensemble of stimulus patterns represented by the configuration of a subset of $N_{\sf S}<N$ nodes, ${\cal C}_{\sf S}=\{Z_{\sf S}=(z_{i_1},z_{i_2},\hdots,z_{i_{N_{\sf S}}})|z_{i_k}\in S\}$ with $\dim{\cal C}_{\sf S}=(\tau_{\sf r}+1)^{N_{\sf S}}$, and an ensemble of corresponding response patterns represented by the configuration of a subset of $N_{\sf R}<N$ nodes, ${\cal C}_{\sf R}=\{Z_{\sf R}=(z_{j_1},z_{j_2},\hdots,z_{j_{N_{\sf R}}})|z_{j_m}\in S\}$ with $\dim{\cal C}_{\sf R}=(\tau_{\sf r}+1)^{N_{\sf R}}$, where $i_k$ and $j_m$ belong to random, disjoint subsets (of dimensions $N_{\sf S}$ and $N_{\sf R}$, respectively) of the set of all $N$ nodes. We thus have \cite{ShannonWeaver1948}: $I_T({\sf S};{\sf R}) = H({\sf R}) - H({\sf R}|{\sf S})$, where $H({\sf R}) = -\sum_{{\cal C}_{\sf R}} P(Z_{\sf R})\log_2 P(Z_{\sf R})$ is the entropy (or variability) of the responses with $P(Z_{\sf R})=N_{Z_{\sf R}}/{(\tau_{\sf r}+1)^{N_{\sf R}} N_{\sf trials}}$, and $H({\sf R}|{\sf S}) = -\sum_{{\cal C}_{\sf R},{\cal C}_{\sf S}} P(Z_{\sf R}|Z_{\sf S})\log_2 P(Z_{\sf R}|Z_{\sf S})$ is the entropy of the responses conditioned on the stimuli with $P(Z_{{\sf R}}|Z_{{\sf S}})=N_{{Z_{\sf R}|Z_{\sf S}}}/N_{\sf trials}$. In the equations above, $N_{Z_{\sf R}}$ corresponds to the number of times the configuration $Z_{\sf R}$ appears in the response and $N_{{Z_{\sf R}|Z_{\sf S}}}$ corresponds to the number of times the configuration $Z_{\sf R}$ appears in response only to the stimulus $Z_{\sf S}$. The subscript $T$ in the mutual information is an integer representing the number of time steps between the stimulus and the response.

\begin{figure}
	\includegraphics[width=\columnwidth]{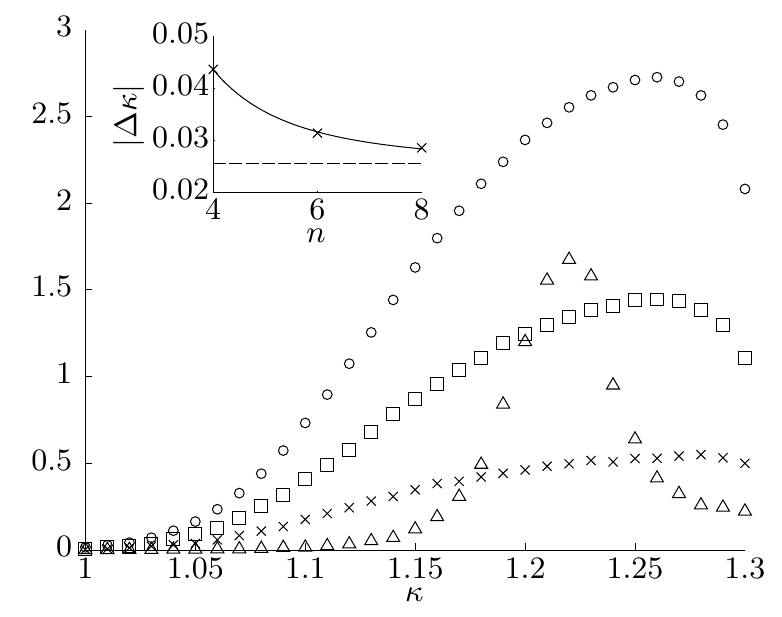}
	\caption{Dynamical susceptibility $\chi(\kappa)$ (triangles) and average mutual information $I(\kappa)$ (in bits). The mutual information is computed for values $n=\{4,6,8\}$, shown as crosses, squares, and circles, respectively (main figure). The discrepancy between the Widom line and average mutual information peaks, $|\Delta \kappa|$, is determined for each value of $n$ (crosses) along with the line of best-fit (solid line: $|\Delta \kappa|=a n^{-b}+c$) which approaches $0.026$ (dashed line) as $n$ is increased (inset).}
	\label{MIpeaks}
\end{figure}

We set $N_{\sf S}=N_{\sf R}=n$ and start a CBM simulation with an initial network configuration corresponding to an element of the stimulus configuration ensemble ${\cal C}_{{\sf S}}$; the resulting mutual information is computed using the configuration of the response nodes after some delay, i.e. some number of time steps $T$ later. The average mutual information at a particular value of the branching ratio $I(\kappa)$ is determined after each element of the stimulus node configuration ensemble ${\cal C}_{\sf S}$ has been repeatedly applied $N_{\sf trials}$ times and averaged over the set of $T=\{T_{\sf min},T_{\sf min}+\delta T,\hdots,T_{\sf max} \}$ delay times, i.e.
\begin{equation}
	I(\kappa) = \frac{1}{N_{\sf delays}}\sum_{T=T_{\sf min}}^{T_{\sf max}} I_T({\sf S};{\sf R}) ,
\end{equation}
where $\delta T=T_{\sf min}+(T_{\sf max}-T_{\sf min})/(N_{\sf delays}-1)$. Clearly, the task of computing $I(\kappa)$ quickly becomes numerically-intensive as $n$ is increased. Using a system size of $N=64$ and $N_{\sf trials}=100$, we compute the mutual information for different sizes of input/output node sets $n=\{4,6,8\}$ averaged over the delays $T=\{60,...,120\}$ in steps of $\delta T=10$ (so $T_{\sf min}=60$, $T_{\sf max}=120$, and $N_{\sf delays}=7$) to demonstrate that the peak in $I(\kappa)$ converges towards the Widom line, i.e. the peak in the dynamical susceptibility at $\kappa_{\sf w}\approx 1.22$, as $n$ is increased (see Fig. \ref{MIpeaks}). Peak locations for $I(\kappa)$ were determined by fitting to third-order polynomials and identifying $\kappa$ values which corresponded to the maxima. As $n$ approaches $N/2$ in the thermodynamic limit, we expect the $I(\kappa)$ and $\chi(\kappa)$ peaks to precisely overlap. We note that whereas mutual information has previously been shown to peak at the location of a phase transition in a variety of systems \cite{Matsuda1996,Wicks2007}, we argue based on our numerical evidence
 that generally the mutual information peaks along the Widom line.

\section{Conclusions and Outlook}
The central purpose of this article is the rejection of the criticality hypothesis and the introduction of a novel, quasi-critical framework to take its place. To this end, we have introduced the cortical branching model (CBM), determined its non-equilibrium phase diagram, and developed a mean-field approximation. One can distinguish the following non-equilibrium phases: a subcritical disordered phase, a supercritical ordered phase, and an oscillatory quasiperiodic phase. In our CBM, we have identified four timescales, three of which we manipulate here: (1) the driving timescale associated with the spontaneous activation probability $p_{\sf s}$, (2) the relaxation timescale associated with the branching parameter $\kappa$, (3) the refractory timescale associated with the refractory period $\tau_{\sf r}$, and (4) the transmission timescale, i.e. the time a signal is in transit from its origin to its destination node. Indeed, the existence of multiple timescales is characteristic of self-organized critical (SOC) phenomena, although in that case, the driving and relaxation timescales are the only typically relevant ones.

Key to our main quasi-criticality hypothesis is the concept of a non-equilibrium Widom line, a line of maximal dynamical susceptibility, which naturally leads to a set of specific questions which can be addressed in living neural networks. For example: What is the location and extent of the non-equilibrium Widom line in the space of $(p_{\sf s}, \kappa, \tau_{\sf r})$? By what factor is the maximum susceptibility modified by changes in $p_{\sf s}$? Most importantly, what mechanisms drive living neural networks towards our non-equilibrium Widom line? All of these questions are experimentally accessible because manipulations of $p_{\sf s}$, $\kappa$, and $\tau_{\sf r}$ are readily realized with the perfusion of pharmacological agents, adjustments of ionic concentrations \cite{Shew2011,Chiappalone2003}, or the control of background stimulation \cite{Vajda2008,Gunning2013,Wagenaar2005}. We remark that in changing $\tau_{\sf r}$, we are manipulating an intrinsic timescale of the system. There are a number of ways living neural networks could adjust such a timescale: as a result of widespread neuronal activation or synchronization, or perhaps by changing the balance of excitation and inhibition.

This novel framework may also serve to explain existing experimental results. For instance, although there have been numerous reports of power laws resulting from spiking activity \emph{in vitro} \cite{Mazzoni2007,Pasquale2008,Friedman2012}, they are rarely found \emph{in vivo} \cite{Hahn2010,Ribeiro2010}. In the context of what is presented here, \emph{in vitro} preparations could have a much smaller $p_{\sf s}$ than \emph{in vivo} preparations, which would suggest that they operate closer to criticality. And although the influence of different spontaneous activation probability distributions (e.g. Poisson, geometric, naturalistic) on the phase diagram or on details of the Widom line is not explored here, it could be probed experimentally to answer questions relating to the effect of external stimuli on the brain. Isolated neural networks used for \emph{in vitro} preparations typically show intervals of many seconds between network bursts that initiate neuronal avalanches, while the avalanches themselves last tens to hundreds of milliseconds \cite{Lombardi2012}. This separation of timescales, which is often given as a requirement for SOC \cite{Jensen1998}, is not clearly seen with \emph{in vivo} preparations, where each neural network receives many synaptic inputs from other intact brain regions.

The significance of the unveiled quasiperiodic phase in terms of the behavior of living neural networks has not yet been fully explored. Previous studies have found neuronal refractory periods to increase as a result of the axonal demyelination associated with multiple sclerosis \cite{QuandtDavis1992,Felts1997,Luo2013}--a disease which is correlated with an unexplained increased incidence of epileptic seizures \cite{KelleyRodriguez2009}. Perfusion of glutamate receptor agonists (such as kainic acid, KA) has been found to decrease neuronal refractory periods, while glutamate receptor antagonists (such as 6-cyano-7-nitroquinoxaline-2,3-dione, CNQX) were found to increase them \cite{DawkinsSewell2004}. Paradoxically, both KA and CNQX have been used to induce {\it in vitro} seizure-like activity \cite{Bragin1999,vanDrongelen2005}. So while the oscillations observed in simulations are possibly related to the pathological synchronization typically associated with epilepsy, we note that synchronization in epilepsy is complex and not yet fully understood \cite{Jiruska2013}.

Finally from a general theoretical standpoint, we would like to state that the influence and importance of network topology has not escaped our notice. In this article, we have only used irreducible random directed graphs with fixed in-degree, partly to facilitate the development of the mean-field approximation presented herein. It would be interesting to explore other network topologies, including reducible and non-planar directed graphs, and additionally study numerically and, if possible, develop mean-field-like approximations of what may lead to an entirely different paradigm of non-equilibrium dynamics.

\section*{Acknowledgments}
The authors would like to thank J. Kert\'{e}sz, K.M. Pilgrim, and A. Vespignani for helpful discussions.

\end{document}